\documentclass{article}
\usepackage{spconf,amsmath,graphicx,hyperref}
\pdfpagewidth=\paperwidth
\pdfpageheight=\paperheight

\usepackage{times}
\usepackage{latexsym}
\usepackage{multirow}
\usepackage[T1]{fontenc}
\usepackage[utf8]{inputenc}
\usepackage{booktabs}
\usepackage{tabularx}
\usepackage{array}
\usepackage[table]{xcolor}
\usepackage{pifont}
\usepackage{microtype}
\hypersetup{hidelinks}
\IfFileExists{inconsolata.sty}{\usepackage{inconsolata}}{}

\newcolumntype{Y}{>{\raggedright\arraybackslash}X}
\newcommand{\cmark}{\ding{51}}
\newcommand{\xmark}{\ding{55}}
\newcommand{\correct}[1]{\cellcolor{green!10}{\textcolor{green!45!black}{\textbf{\cmark}}~#1}}
\newcommand{\wrong}[1]{\cellcolor{red!8}{\textcolor{red!65!black}{\textbf{\xmark}}~#1}}

\title{Robustness of Text-Only Adapted Speech LLMs \\ for Spoken Semantic Parsing}
\name{\normalsize
    Youngwon Choi\textsuperscript{1,*},
    Donghyuk Jung\textsuperscript{2,*},
    Hyeonyu Kim\textsuperscript{1},
    Taeyoun Kwon\textsuperscript{1,3},
    Myeongkyun Cho\textsuperscript{1,4},
    Moon Hwan Kim\textsuperscript{1,5}
}
\address{\normalsize
    \textsuperscript{1}Maum AI Inc.,
    \textsuperscript{2}Korea Culture Technology Institute,
    \textsuperscript{3}Seoul National University,
    \textsuperscript{4}KAIST,
    \textsuperscript{5}Yonsei University
    \\
    \small\textsuperscript{*}These authors contributed equally to this work.
}

\begin{document}
%
\maketitle
\begin{abstract}
Large language models (LLMs) are increasingly used to generate structured outputs from natural language instructions, including semantic frames and function calls. 
When the input is speech, this task becomes spoken semantic parsing, which can be performed using either Speech LLMs or ASR–LLM cascades. 
Although both approaches can be adapted with text-only task supervision, their relative robustness to acoustic degradation remains unclear. 
We compare these approaches under matched text-form supervision using 3B and 7B models across three datasets, evaluated on clean speech and babble noise at five SNR levels.
Results show that Speech LLMs generally achieve better performance in command understanding, intent prediction, and function calling, particularly under severe noise, while improvements in fine-grained entity prediction are less consistent.
An additional evaluation shows that Speech LLMs achieve lower average inference time per sample than ASR--LLM cascades at both model scales.
\end{abstract}
\section{Introduction}
\label{sec:intro}

Large language models (LLMs) are increasingly used to generate structured outputs from natural language instructions, including semantic representations~\cite{mekala2023zerotop}, executable actions~\cite{huang2022language}, and tool calls~\cite{schick2023toolformer}, providing machine-readable interfaces for downstream systems.
Such capabilities can be elicited through prompting and in-context reasoning~\cite{yao2023react, lu2023chameleon}, while supervised fine-tuning (SFT) on text-form input--output pairs offers an alternative for adapting models to task-specific schemas and output formats~\cite{qin2024toolllm, patil2024gorilla}.
When such mappings are performed from spoken rather than written instructions, the task becomes spoken semantic parsing, requiring models to recover semantic information from speech and map it to structured representations.

For spoken semantic parsing, two paradigms are commonly used.
An ASR--LLM cascade first transcribes speech into text using an automatic speech recognition (ASR) system and then passes the transcript to an LLM to generate structured outputs \cite{everson2024towards,huang2024audiogpt}.
Although modular and easy to integrate, this approach is susceptible to error propagation, as recognition errors can distort the semantic information required for accurate structured predictions \cite{faruqui2022revisiting,shapira2025measuring,jung2026analyzing}.
Alternatively, Speech LLMs~\cite{chu2024qwen2,tang2024salmonn,xu2025qwen25omni,fathullah2024audiochatllama}, which can directly process speech and generate text-based outputs, can perform semantic parsing without an explicit transcription step~\cite{hacioglu2025speechllms, choi2025desamo, pahwa2026audio2tool}.

In practical settings, adapting these systems to downstream tasks does not necessarily require paired speech--target annotations, which can be costly to collect for each target task \cite{qin2021survey,huang2020leveraging}.
For ASR--LLM cascades, the downstream LLM can be naturally fine-tuned using text-form input--output pairs.

Similarly, Speech LLMs can be adapted through text-only fine-tuning of their language model components, leveraging the learned correspondence between speech and text~\cite{xiang2025understanding,kim2026spoken} to transfer task knowledge to speech inputs at inference time, an approach explored in both ASR and spoken language understanding~\cite{ma2024effective,fang2025low,choi2026exploring}.
However, the acoustic robustness of text-only adapted Speech LLMs has not been systematically compared with that of ASR--LLM cascades under matched text-form supervision.

In this work, we conduct a systematic comparison of text-only adapted Speech LLMs and ASR--LLM cascades for spoken semantic parsing under acoustic degradation.
We evaluate both approaches using matched language model scales across three datasets under babble noise at multiple SNR levels.
Our results show that Speech LLMs generally exhibit greater robustness in high-level semantic prediction, although this advantage does not consistently extend to fine-grained entity grounding.
We further find that text-only adapted Speech LLMs require less average inference time per sample than scale-matched ASR--LLM cascades at both model scales.

\begin{table*}[t]
\centering
\small
\begin{tabular}{l l l r r r}
\toprule
Dataset & Target Format & Metric & \# Train & \# Dev & \# Test \\
\midrule
FSC 
& \texttt{\{action, object, location\}} 
& EM 
& 23132 & 3118 & 3793 \\

SLURP 
& \texttt{\{scenario, action, entities\}} 
& Intent Acc., Entity F1, SLU-F1 
& 11514 & 2033 & 13078 \\

In-house 
& Function call with arguments 
& EM 
& 21397 & 891 & 591 \\
\bottomrule
\end{tabular}
\caption{Overview of the datasets used in our experiments. Target formats are JSON-style semantic frames for FSC and SLURP, and function calls for the in-house dataset. Train and development splits are used only as text-form supervision. For SLURP, train and development instances are constructed at the sentence level, while test instances correspond to audio-level utterances.}
\label{tab:datasets}
\end{table*}

\section{Experimental Setup}
\label{sec:setup}

\subsection{Text-Only Adaptation of the Compared Systems}
\label{sec:adaptation}

Let $a_i$ be a spoken user query, $x_i$ its text form, and $y_i$ the target structured output (an intent with slot values, a semantic frame, or a function call), serialized as a text sequence.
Both systems are adapted from text-only task supervision $\mathcal{D}_{\text{text}}=\{(x_i,y_i)\}_{i=1}^{N}$; no speech--target pairs $(a_i,y_i)$ are used during adaptation.
Each example is placed into a task prompt template $T(\cdot)$ that contains the task instruction and a query slot, and the model is fine-tuned with the next-token prediction objective
\begin{equation}
\mathcal{L}(\theta) = -\sum_{i=1}^{N} \log p_{\theta}\left(y_i \mid T(x_i)\right),
\label{eq:loss}
\end{equation}
where $\theta$ denotes the trainable parameters.

For the Speech LLM, $\theta$ comprises only the parameters of its LLM component; the speech encoder and modality adapter, which receive no supervision from text-only training, are kept frozen.
At inference, the spoken query $a_i$ is inserted through an audio placeholder into the same query slot that $x_i$ occupied during training, and the structured output is decoded directly from speech,
\begin{equation}
\hat{y}_i = \mathrm{Decode}\left(p_{\theta}(\cdot \mid T(a_i))\right),
\label{eq:decode}
\end{equation}
relying on pretrained speech--text alignment to transfer the learned task mapping to speech inputs.
For the ASR--LLM cascade, the ASR system first generates a transcript $\hat{x}_i$ from $a_i$, which is then provided to the fine-tuned LLM:
\begin{equation}
\hat{y}_i = \mathrm{Decode}\left(p_{\theta}(\cdot \mid T(\hat{x}_i))\right).
\end{equation}

\subsection{Datasets and Noise Conditions}
\label{sec:data}

We evaluate on three datasets: Fluent Speech Commands (FSC)~\cite{lugosch2019speech}, SLURP~\cite{bastianelli2020slurp}, and an in-house product dataset.
FSC is a smart-home command understanding dataset with action, object, and location annotations, and is evaluated using exact match (EM) over the full semantic frame.
SLURP is a spoken language understanding dataset of single-turn home-assistant interactions with scenario, action, and entity annotations; compared with FSC, it covers broader linguistic and semantic variation.
We evaluate SLURP using intent accuracy, entity F1, and SLU-F1, following the standard SLURP evaluation protocol.
The in-house dataset is an English smart-home function calling dataset for product-oriented voice agent commands.
Inspired by Octopus v2~\cite{chen2024octopus}, each command is paired with one of 42 predefined target functions and its arguments, but we do not use special function tokens or function descriptions in the prompt.
The speech data were recorded in the target product environment under clean acoustic conditions and include the product-level audio preprocessing used by the voice agent system.
We evaluate this dataset using EM over the generated function call and arguments.
Table~\ref{tab:datasets} summarizes the target format, evaluation metrics, and the number of examples in the train, development, and test splits.

For noisy evaluation, we generate babble noise using the speech subset of the MUSAN corpus~\cite{snyder2015musan}, following the augmentation procedure adopted in the Kaldi speaker-recognition recipes~\cite{snyder2018x}.
Specifically, we mix three to seven randomly selected speech segments to create background babble noise, and add it to each test utterance at SNR levels of 15, 10, 5, 2.5, and 0 dB.

\begin{table*}[t]
\centering
\small
\setlength{\tabcolsep}{5pt}
\renewcommand{\arraystretch}{1.0}
\begin{tabular}{ll l l c cccccc}
\toprule
\multirow{2}{*}{Dataset} & \multirow{2}{*}{Metric} & \multirow{2}{*}{Scale} & \multirow{2}{*}{System} & \multirow{2}{*}{Oracle} & \multicolumn{6}{c}{Speech input condition} \\
\cmidrule(lr){6-11}
& & & & & Clean & 15 dB & 10 dB & 5 dB & 2.5 dB & 0 dB \\
\midrule
\multirow{4}{*}{FSC} & \multirow{4}{*}{EM}
  & \multirow{2}{*}{3B} & SLM     & 100.00 & \textbf{98.63} & \textbf{96.78} & \textbf{93.86} & \textbf{85.26} & \textbf{76.43} & \textbf{63.14} \\
& &                     & Cascade & 100.00 & 98.21 & 95.49 & 91.77 & 82.47 & 73.56 & 59.45 \\
& & \multirow{2}{*}{7B} & SLM     & 100.00 & \textbf{99.26} & \textbf{97.73} & \textbf{95.25} & \textbf{87.82} & \textbf{79.07} & \textbf{65.81} \\
& &                     & Cascade & 100.00 & 98.26 & 95.52 & 91.93 & 82.86 & 74.43 & 60.22 \\
\midrule
\multirow{12}{*}{SLURP} & \multirow{4}{*}{Intent Acc.}
  & \multirow{2}{*}{3B} & SLM     & 90.71 & \textbf{84.11} & \textbf{79.32} & \textbf{73.08} & \textbf{57.43} & \textbf{46.13} & \textbf{32.53} \\
& &                     & Cascade & 89.88 & 81.27 & 75.19 & 67.21 & 49.90 & 38.65 & 26.19 \\
& & \multirow{2}{*}{7B} & SLM     & 91.57 & \textbf{86.02} & \textbf{81.34} & \textbf{74.94} & \textbf{59.47} & \textbf{46.68} & \textbf{33.29} \\
& &                     & Cascade & 91.21 & 81.97 & 75.75 & 67.98 & 50.69 & 38.91 & 26.14 \\
\cmidrule(lr){2-11}
& \multirow{4}{*}{Entity F1}
  & \multirow{2}{*}{3B} & SLM     & 82.30 & \textbf{61.36} & \textbf{56.26} & \textbf{50.84} & 38.78 & 30.27 & 20.21 \\
& &                     & Cascade & 82.16 & 59.64 & 55.64 & 50.55 & \textbf{39.04} & \textbf{30.49} & \textbf{20.44} \\
& & \multirow{2}{*}{7B} & SLM     & 83.17 & \textbf{65.11} & \textbf{60.45} & \textbf{54.54} & \textbf{42.29} & \textbf{32.62} & \textbf{22.39} \\
& &                     & Cascade & 83.86 & 59.84 & 56.05 & 51.08 & 39.63 & 31.59 & 20.68 \\
\cmidrule(lr){2-11}
& \multirow{4}{*}{SLU-F1}
  & \multirow{2}{*}{3B} & SLM     & 84.60 & \textbf{70.64} & 66.23 & 61.25 & 49.35 & 40.38 & 28.74 \\
& &                     & Cascade & 84.73 & 70.34 & \textbf{66.90} & \textbf{62.12} & \textbf{50.81} & \textbf{42.90} & \textbf{29.74} \\
& & \multirow{2}{*}{7B} & SLM     & 84.92 & \textbf{73.77} & \textbf{69.86} & \textbf{64.40} & \textbf{52.86} & \textbf{47.66} & \textbf{31.21} \\
& &                     & Cascade & 86.01 & 70.23 & 66.98 & 62.41 & 51.36 & 43.26 & 29.97 \\
\midrule
\multirow{4}{*}{In-house} & \multirow{4}{*}{EM}
  & \multirow{2}{*}{3B} & SLM     & 95.94 & 89.51 & 88.32 & 84.77 & \textbf{75.30} & \textbf{64.30} & \textbf{49.24} \\
& &                     & Cascade & 95.43 & \textbf{90.86} & \textbf{89.00} & \textbf{85.28} & 72.59 & 60.07 & 44.67 \\
& & \multirow{2}{*}{7B} & SLM     & 94.42 & \textbf{90.19} & \textbf{89.68} & \textbf{86.80} & \textbf{76.82} & \textbf{67.34} & \textbf{55.33} \\
& &                     & Cascade & 93.91 & 89.68 & 86.97 & 83.93 & 73.60 & 62.10 & 44.84 \\
\bottomrule
\end{tabular}
\caption{Main results of the text-only adapted Speech LLM and the ASR--LLM cascade under babble noise. FSC and the in-house dataset report exact match (EM); SLURP reports intent accuracy, entity F1, and SLU-F1. Oracle denotes evaluation using gold transcripts; dB values denote the SNR of the noisy speech input. The better of the two systems for each dataset, metric, scale, and condition is in bold.}
\label{tab:main_results}
\end{table*}

\subsection{Models and Compared Systems}
\label{sec:systems}

We compare text-only adapted Speech LLMs against an ASR--LLM cascade of comparable scale.
For Speech LLMs, we use the Thinker components of Qwen2.5-Omni-3B and Qwen2.5-Omni-7B~\cite{xu2025qwen25omni} as our 3B and 7B backbones, respectively.
These models are fine-tuned with text-form task supervision and directly receive speech input at inference time.

For the ASR--LLM cascade baseline, we use Whisper large-v3~\cite{radford2023robust} as the ASR module and Qwen2.5-3B-Instruct or Qwen2.5-7B-Instruct~\cite{yang2024qwen25} as the downstream LLM.
Each cascade is paired with the corresponding Speech LLM configuration: Whisper large-v3 with Qwen2.5-3B-Instruct is compared against Qwen2.5-Omni-3B, while Whisper large-v3 with Qwen2.5-7B-Instruct is compared against Qwen2.5-Omni-7B.
The downstream LLMs in the cascades are fine-tuned with the same text-form supervision used for Speech LLMs.
This setup allows us to compare direct speech-based inference with transcript-based inference while keeping the task adaptation data consistent across systems.

We also include an oracle-text setting, where the fine-tuned models receive gold transcripts at inference time.
This setting provides an upper-bound reference for evaluating structured output generation without transcription errors.

\subsection{Training and Inference Details}
We largely follow the training setup of our previous work~\cite{choi2026exploring}, unless otherwise noted.
Specifically, all trainable language-model components are fine-tuned for three epochs using the AdamW optimizer with bfloat16 precision.
Training is conducted on four NVIDIA A100 GPUs, with a per-device batch size of 2 and 8 gradient accumulation steps.
We use a learning rate of $5.0 \times 10^{-6}$ and a warmup ratio of 0.04. 
At inference, all models use a batch size of 8, with greedy decoding for the LLMs and default decoding for Whisper large-v3.

\section{Results and Analysis}
\label{sec:results}

\subsection{Main Results}
\label{sec:robustness}
Table~\ref{tab:main_results} shows the performance of text-only adapted Speech LLMs and ASR--LLM cascades under clean and noisy speech conditions.
On FSC, text-only adapted Speech LLMs consistently outperform the cascades across all speech conditions and both model scales.
For example, with the 7B backbone, the absolute EM gain increases from 1.00 under clean speech to 5.59 at 0 dB.

On SLURP, Speech LLMs show the clearest gains in intent accuracy, suggesting that direct speech-based inference can be advantageous for high-level semantic prediction.
With the 7B backbone, Speech LLMs improve intent accuracy over the cascades by 4.05, 8.78, and 7.15 percentage points under clean speech, 5 dB, and 0 dB, respectively.
The gains on entity F1 and SLU-F1 are more mixed, especially with the 3B backbone, suggesting that text-only adapted Speech LLMs are more reliable for intent-level prediction than for fine-grained entity prediction.

On the in-house function-calling dataset, Speech LLMs generally show greater advantages under more severe noise conditions.
In the 3B setting, the cascade is slightly better under clean and mild-noise conditions, but Speech LLMs surpass it at 5 dB and below, reaching an absolute EM gain of 4.57 at 0 dB.
In the 7B setting, Speech LLMs outperform the cascades across all speech conditions, with the largest gain at 0 dB, improving EM from 44.84 to 55.33.

Taken together, these results suggest that text-only adapted Speech LLMs can offer advantages over ASR--LLM cascades in high-level semantic prediction, particularly under challenging noise conditions, although improvements in fine-grained entity prediction are less consistent.

\begin{table*}[t]
\centering
\small
\setlength{\tabcolsep}{2.5pt}
\renewcommand{\arraystretch}{1.1}
\begin{tabularx}{\textwidth}{>{\raggedright\arraybackslash}p{1.25cm} >{\raggedright\arraybackslash}p{2.4cm} >{\raggedright\arraybackslash}p{2.6cm} Y Y}
\toprule
Dataset & Gold Transcript & ASR Hypothesis & ASR--LLM Cascade & Text-Only Adapted SLM \\
\midrule

FSC
&
Heat down
&
Feet down.
&
\wrong{\texttt{\{action: deactivate, object: feet, location: none\}}}
&
\correct{\texttt{\{action: decrease, object: heat, location: none\}}}
\\

\midrule

SLURP
&
is it going to rain on monday
&
Is it going straight on Monday?
&
\wrong{\texttt{\{scenario: transport, action: query, entities: [date: Monday]\}}}
&
\correct{\texttt{\{scenario: weather, action: query, entities: [weather\_descriptor: rain, date: monday]\}}}
\\

\midrule

In-house
&
Turn off the screen brightness.
&
turn off this room privately.
&
\wrong{\texttt{set\_privacy\_mode(\allowbreak enable=False)}}
&
\correct{\texttt{set\_brightness(\allowbreak brightness=0)}}
\\

\midrule

SLURP
&
i would like to hear some rap music
&
I would like to hear some rap music.
&
\correct{\texttt{\{scenario: play, action: music, entities: [music\_genre: rap]\}}}
&
\wrong{\texttt{\{scenario: play, action: music, entities: [music\_genre: rock]\}}}
\\

\bottomrule
\end{tabularx}
\caption{Qualitative examples comparing the ASR--LLM cascade and the text-only adapted Speech LLM at 5 dB SNR. The Speech LLM uses Qwen2.5-Omni-3B, while the cascade uses Whisper large-v3 and Qwen2.5-3B-Instruct. Green check marks indicate outputs that match the gold annotation, while red crosses indicate incorrect outputs.}
\label{tab:qualitative_examples}
\end{table*}

\begin{table}[t]
\centering
\small
\setlength{\tabcolsep}{5pt}
\renewcommand{\arraystretch}{1.1}
\begin{tabular}{lcc}
\toprule
Scale & Speech LLM & ASR--LLM Cascade \\
& (sec/sample) & (sec/sample) \\
\midrule
3B & \textbf{1.072} & 1.304 \\
7B & \textbf{0.882} & 1.108 \\
\bottomrule
\end{tabular}
\caption{Average end-to-end inference time per sample of text-only adapted Speech LLMs and ASR--LLM cascades across model scales.
For cascades, the reported time includes both ASR transcription and LLM generation.}
\label{tab:latency}
\end{table}

\subsection{Qualitative Analysis}

Table~\ref{tab:qualitative_examples} presents qualitative examples comparing the ASR--LLM cascade and text-only adapted Speech LLM.
The first three examples illustrate how ASR errors can distort task-relevant words and propagate to downstream semantic predictions.
For example, ``Heat down'' is transcribed as ``Feet down,'' causing the cascade to predict the wrong object and action.
In contrast, the Speech LLM can often recover the intended task semantics directly from speech, even when the corresponding ASR transcript is misleading or severely corrupted.

The last example shows the failure mode of Speech LLMs.
The Speech LLM correctly identifies the high-level scenario and action, but replaces the fine-grained entity value ``rap'' with ``rock.''
This suggests that direct speech-based prediction can capture high-level task semantics but may still struggle with exact lexical grounding for slot values.
In contrast, when the ASR transcript is correct, the cascade can benefit from the explicit textual intermediate and recover the correct entity.

\subsection{Inference Time and Deployment Considerations}

We additionally measure average end-to-end inference time per sample on SLURP utterances using a single NVIDIA A100 GPU in the same Hugging Face Transformers-based inference environment.
As shown in Table~\ref{tab:latency}, text-only adapted Speech LLMs require less average inference time per sample than scale-matched ASR--LLM cascades at both model scales.
Specifically, Speech LLMs reduce average inference time per sample by 17.8\% and 20.4\% for the 3B and 7B configurations, respectively.
This reduction may partly reflect the additional ASR decoding stage required by cascaded systems.
Interestingly, the 3B configurations have higher average inference times per sample than their 7B counterparts, potentially due to their greater decoder depth (36 vs. 28 layers).

These results show that text-only adapted Speech LLMs require less average end-to-end inference time per sample than scale-matched ASR--LLM cascades at both model scales.
Combined with their robustness in high-level semantic prediction, this suggests that direct speech-based inference can be advantageous when inference efficiency and semantic understanding under acoustic degradation are important.

\section{Conclusion}
\label{sec:conclusion}

We compared text-only adapted Speech LLMs and ASR--LLM cascades for spoken semantic parsing under identical text-form supervision across three datasets and two model scales.
Results show that Speech LLMs generally exhibit greater robustness in high-level semantic prediction, although their advantages in fine-grained entity prediction are less consistent and depend on the task and model scale.
The two approaches also exhibit distinct failure patterns: cascades propagate ASR errors, whereas Speech LLMs may recover the intended semantics but produce incorrect slot values.
Additionally, Speech LLMs achieve lower average inference time per sample than scale-matched cascades.
However, our evaluation covers only one model family and uses a single training seed, which may limit the generalizability of our findings.

\section{Acknowledgments}
This research was supported by Culture, Sports and Tourism R\&D Program through the Korea Creative Content Agency grant funded by the Ministry of Culture, Sports and Tourism in 2026 (Project Name: Development of AI-based personalized cultural and arts learning services using smart device, Project Number: RS-2026-25524629). OpenAI's ChatGPT was used to assist with manuscript writing and experimental code development.
All AI-assisted content and code were reviewed and verified by the authors.

\section{Compliance with Ethical Standards}
The in-house speech data were collected for internal voice-agent development, with permission for research and evaluation.
The associated transcripts and text-form training examples contain no personally identifiable information, and the dataset is not publicly released.

\begingroup
\let\originalthebibliography\thebibliography
\renewcommand{\thebibliography}[1]{%
    \originalthebibliography{#1}%
    \fontsize{9pt}{10.5pt}\selectfont%
}
\bibliographystyle{IEEEbib}
\bibliography{refs}
\endgroup

\end{document}